\documentclass[12pt]{article}
\usepackage{fullpage,epsfig,psfrag,graphics,amsbsy,amssymb,citesort}
\usepackage{caption}
\newcommand{\beq}{\begin{equation}}
\newcommand{\eeq}{\end{equation}}
\newcommand{\bea}{\begin{eqnarray}}
\newcommand{\eea}{\end{eqnarray}}
\newcommand{\bfs}{\boldsymbol}

\newcommand{\be}{\begin{equation}}
\newcommand{\ee}{\end{equation}}
\newcommand{\bq}{\begin{eqnarray}}
\newcommand{\eq}{\end{eqnarray}}
\newcommand{\ket}[1]{|#1\rangle}
\newcommand{\bra}[1]{\langle#1|}

\def\math{\mathsurround=0pt }
\def\leftrightarrowfill{$\math \mathord\leftarrow \mkern-6mu 
 \cleaders\hbox{$\mkern-2mu \mathord- \mkern-2mu$}\hfill
 \mkern-6mu \mathord\rightarrow$}
\def\overleftrightarrow#1{\vbox{\ialign{##\crcr
     \leftrightarrowfill\crcr\noalign{\kern-1pt\nointerlineskip}
     $\hfil\displaystyle{#1}\hfil$\crcr}}}

\let\l=\lambda

 \def\bd{\begin{document}} \def\ed{\end{document}}
\def\ds{\documentstyle} \let\fr=\frac \let\bl=\bigl \let\br=\bigr
\let\Br=\Bigr \let\Bl=\Bigl
\let\bm=\bibitem
\let\na=\nabla
\let\pa=\partial \let\ov=\overline
\def\ft#1#2{{\textstyle{{\scriptstyle #1}\over {\scriptstyle #2}}}}
\def\fft#1#2{{#1 \over #2}}
\def\vp{\varphi}
\def\sst#1{{\scriptscriptstyle #1}}
\def\oneone{\rlap 1\mkern4mu{\rm l}}
\def\td{\tilde}
\def\wtd{\widetilde}
\def\dalemb#1#2{{\vbox{\hrule height .#2pt
        \hbox{\vrule width.#2pt height#1pt \kern#1pt
                \vrule width.#2pt}
        \hrule height.#2pt}}}
\def\square{\mathord{\dalemb{6.8}{7}\hbox{\hskip1pt}}}
\def\wtd{\widetilde}
\def\R{\rlap{\rm I}\mkern3mu{\rm R}}
\def\im{{\rm i}}
\def\tilg{\tilde{g}}
\def\tilF{\tilde{F}}
\def\tilA{\tilde{A}}
\def\varf{\varphi}
\def\tilf{\tilde{\phi}}
\def\tilh{\tilde{h}}
\def\rme{{\rm e}}
\def\ep{\epsilon}
\def\0{{(0)}}
\def\9{{(9)}}
\def\8{{(8)}}
\def\7{{(7)}}
\def\6{{(6)}}
\def\5{{(5)}}
\def\4{{(4)}}
\def\3{{(3)}}
\def\2{{(2)}}
\def\1{{(1)}}
\newcommand{\trace}{{\rm Tr}}
\newcommand{\ub}{\overline{U}}
\newcommand{\vb}{\overline{V}}
\newcommand{\uh}{\widehat{U}}
\newcommand{\vh}{\widehat{V}}
\newcommand{\ubh}{\overline{\widehat{U}}}
\newcommand{\vbh}{\overline{\widehat{V}}}
\newcommand{\lb}{\bar{\l}}
\newcommand{\Fb}{\overline{F}}
\newcommand{\Fh}{\widehat{F}}
\newcommand{\Fbh}{\overline{\widehat{F}}}
\newcommand{\Ab}{\overline{A}}
\newcommand{\Ah}{\widehat{A}}
\newcommand{\Abh}{\overline{\widehat{A}}}
\newcommand{\Gb}{\overline{G}}
\newcommand{\Gh}{\widehat{G}}
\newcommand{\Gbh}{\overline{\widehat{G}}}
\newcommand{\Pb}{\overline{P}}
\newcommand{\Ph}{\widehat{P}}
\newcommand{\Pbh}{\overline{\widehat{P}}}
\newcommand{\Qb}{\overline{Q}}
\newcommand{\Qh}{\widehat{Q}}
\newcommand{\Qbh}{\overline{\widehat{Q}}}
\newcommand{\Bb}{\overline{B}}
\newcommand{\Bh}{\widehat{B}}
\newcommand{\Bbh}{\overline{\widehat{B}}}
\newcommand{\fhns}{\hat{F}^{\rm (NS)}}
\newcommand{\fhrr}{\hat{F}^{\rm (RR)}}
\newcommand{\ahns}{\hat{A}^{\rm (NS)}}
\newcommand{\ahrr}{\hat{A}^{\rm (RR)}}
\newcommand{\hhrr}{\hat{H}^{\rm (RR)}}
\newcommand{\hchi}{\hat{\chi}}
\newcommand{\hphi}{\hat{\phi}}
\newcommand{\htau}{\hat{\tau}}
\newcommand{\cG}{{\cal G}}
\newcommand{\cGb}{\overline{{\cal G}}}
\newcommand{\cH}{{\cal H}}
\newcommand{\cP}{{\cal P}}
\newcommand{\cPb}{\overline{{\cal P}}}
\newcommand{\cQ}{{\cal Q}}
\newcommand{\cQb}{\overline{{\cal Q}}}
\newcommand{\cM}{{\cal M}}
\newcommand{\cN}{{\cal N}}
\newcommand{\cO}{{\cal O}}
\newcommand{\cD}{{\cal D}}
\newcommand{\cL}{{\cal L}}

\newcommand{\vpp}{\mbox{$\langle{\scriptstyle++}\rangle$}}
\newcommand{\vmp}{\mbox{$\langle{\scriptstyle-+}\rangle$}}
\newcommand{\vppp}{\mbox{$\langle{\scriptstyle+++}\rangle$}}
\newcommand{\vmpp}{\mbox{$\langle{\scriptstyle-++}\rangle$}}
\newcommand{\vpmp}{\mbox{$\langle{\scriptstyle+-+}\rangle$}}

\begin{document}
\setlength{\captionmargin}{36pt}
\begin{titlepage}
\begin{flushright}
\phantom{UFIFT-HEP-10} 
\end{flushright}

\vskip 2cm
\begin{center}
\begin{large}
{\bf Digitizing the Neveu-Schwarz Model on the 
Lightcone Worldsheet}
\end{large}

\vskip 2cm
{\large 
Charles B. Thorn\footnote{E-mail  address: {\tt thorn@phys.ufl.edu}}
}
\vskip0.8cm
{\it Institute for Fundamental Theory\\
Department of Physics, University of Florida,
Gainesville FL 32611}


\vskip 1.0cm
\end{center}

\begin{abstract}
\noindent The purpose of this article is to extend 
the lightcone worldsheet lattice description of string
theory to include the Neveu-Schwarz model. We
model each component of the fermionic worldsheet
field by a critical Ising model. We show that
a simple choice of boundary conditions for the
Ising variables leads to the half integer modes
required by the model. We identify the G-parity
operation within the Ising model and formulate
the procedure for projecting onto the even G-parity
sector. We construct the lattice version of the
three open string vertex, with the necessary
operator insertion at the interaction point.
We sketch a formalism for summing planar open string
multi-loop amplitudes, and we discuss prospects
for numerically summing them. If successful, the methods
described here could provide an alternative
to lattice gauge theory for computations in large
N QCD. 
\end{abstract}
\vfill
\end{titlepage}
\section{Introduction}
Nonabelian gauge theory in 4 spacetime
dimensions can be regarded as the zero slope limit ($\alpha^\prime\to0$)
of open string theory \cite{neveuscherk} in a higher dimension 
(e.g. $D=10$), provided that the open
strings are required to end on D3-branes \cite{dailp}. 
For the conformally invariant ${\cal N}=4$ supersymmetric gauge theory, 
the AdS/CFT correspondence asserts that, after reinterpreting the
planar open string multiloop diagrams as closed string trees, 
the $\alpha^\prime\to0$ limit remains a string theory on the curved background
AdS$_5\times$S$^5$ \cite{maldacenasole}. In this sense ${\cal N}=4$
isn't just the limit of string theory: it {\it is}
a (closed) string theory. Moreover, the conformal invariance of ${\cal N}=4$
means that the gauge coupling is a true parameter, and this closed 
string theory can be analyzed semiclassically 
in the limit of large 't Hooft coupling $Ng^2$. 

For QCD, however, asymptotic freedom precludes such a parametric
semi-classical closed string limit. Even though 't Hooft's 
large $N$ limit \cite{thooftlargen} of QCD  
should still have an interpretation in terms of some kind of
closed string background, the absence of a semiclassical approximation
to find and analyze that background casts doubt on
the practical value of a closed string interpretation in actual calculations. 
We entertain here the possibility that, in this circumstance, it may
be more profitable to forego the equivalence of 
QCD to some, as yet to be discovered, closed string 
theory  and instead to exploit the known connection of large
$N$ QCD to open string theory, keeping  $\alpha^\prime>0$ as a regulator 
until the end of the analysis. The hope is that
some nonperturbative aspects of QCD, such as quark
confinement and the meson spectrum in `t Hooft's large $N$
limit, will be more tractable treated as an
open string theory with $\alpha^\prime>0$ than as a 
quantum field theory, its $\alpha^\prime\to0$ limit.

Several years ago my collaborators and I showed how to represent each
gauge theory planar diagram as a lightcone open-string worldsheet 
path integral \cite{bardakcit}.
This work established a direct formal connection of large $N$ QCD to
an  open string theory with $\alpha^\prime=0$. 
The problem with keeping $\alpha^\prime=0$, however, is that the
UV divergences of quantum field theory are not properly
regulated, so that various counterterms must be bought in, order
by order in the loop expansion,
to cancel UV induced artifacts that violate Lorentz invariance
\cite{chakrabarti1}. 
Keeping $\alpha^\prime>0$ certainly mitigates these problems,
although it remains to be seen whether it completely removes them.
In any case the formalism for planar graph summation \cite{gilest}
is on firmer foundation, and is actually
somewhat easier to implement, with $\alpha^\prime>0$.

The large $N$ limit of gauge theory amounts to summing all the
planar Feynman diagrams of perturbation theory. These diagrams
are the $\alpha^\prime\to0$ limit of the planar open string
multiloop diagrams. Mandelstam \cite{mandelstamlc}
has given a remarkably
simple and intuitive representation of the latter
diagrams as path integrals over the worldsheets of the light-cone
quantized string \cite{goddardgrt}. Each loop in a given 
multiloop diagram is represented as an internal worldsheet
boundary whose beginning describes a breaking string and whose
end describes two strings joining. For each fixed configuration
of these internal boundaries,
the path integral is a Gaussian integral over the transverse
string coordinates and the measure is precisely the natural lattice
measure for the quadratic action. Summing over the number, location,
and lengths of these internal boundaries with the same measure
accomplishes the sum over planar open string diagrams \cite{gilest}.

While there is still some optimism that the large $N$ limit
of the maximal supersymmetric (${\cal N}=4$)
gauge theory might be exactly solvable \cite{maldacenasole}, 
it is doubtful that this will be possible for
the pure gauge theory underlying QCD. However, it might well
turn out that the representation of the sum of planar
diagrams by a lightcone worldsheet lattice system, as
described in the previous paragraph, can be studied on a
computer, perhaps using Monte Carlo algorithms. Such a
numerical attack on QCD via its connection to
open string theory could have strengths and weaknesses
complementary to those of standard lattice gauge theory
simulations.

The basic framework for this approach to summing
open string planar loop diagrams was set up by Giles and me (GT) over three
decades ago \cite{gilest} in the context of bosonic open string theory
in 26 dimensional spacetime. The zero slope limit of
these diagrams (modulo complications from the presence
of the open string tachyon) would be 26 (not 4) dimensional
gauge theory. Recently, I explained how to incorporate
D3-branes into the GT formalism \cite{thorndirichlet}
in order to arrange zero slope limits that were four
dimensional gauge theories, albeit coupled to 22
massless scalars corresponding to vibrations in the 
extra dimensions. There are mechanisms to decouple such
massless scalars (e.g. see \cite{thornnonabelian}), but
the difficulties of the open bosonic string tachyon would remain.

The open string Neveu-Schwarz model \cite{neveuschwarz,neveust}, 
restricted to  even G-parity
open string states (NS+), has no open string tachyon 
\cite{mandelstamgparity,gliozziso}, so
the multiloop planar diagrams of the $D=4$ version of NS+ 
would have a large $N$ 
4 dimensional gauge theory as a clean zero slope limit \cite{thornsubqcd}. 
But since the standard lightcone quantization works only in the
critical dimension \cite{goddardt,goddardrt,mandelstamnsr}, 
it is simpler to use the 10 dimensional version of
the model, employing D3-branes 
to yield a four dimensional gauge theory. The 6 massless scalars
can then be suppressed either by orbifold projections or by using
nonabelian D3 branes \cite{thornnonabelian}.

The purpose of this article is to adapt the GT worldsheet lattice
formalism to the Neveu-Schwarz model. This requires providing
a viable lattice definition of the fermionic worldsheet field
$H^\mu(\sigma,\tau)$ of that model. Each (transverse) component
of $H$ is a two dimensional Majorana spinor field. As always,
putting fermion fields on a lattice involves difficulties.
Besides the inevitable fermion doubling, for which there are a variety
of remedies (Wilson fermions, staggered fermions, domain wall fermions),
there is the difficulty that Grassmann path integrands do not
have a probabilistic interpretation, a prerequisite for Monte Carlo
methods. This last difficulty could be dealt with by integrating
out the fermions, but that would sacrifice locality, rendering
numerical simulations very costly. Fortunately, for the two
dimensional worldsheet, there is another option which we
pursue here. This is based on the well-known fact that the
physics of the critical two dimensional Ising model is
that of a free Majorana Fermi field. The partition function of the
Ising model is a sum over Ising spins $s_{ij}=\pm1$ with a positive definite 
Boltzmann factor, the exponent of which is local in the spins.
By replacing each component of $H$ with an Ising spin system, the
lightcone lattice formalism for summing the planar
diagrams of the Neveu-Schwarz model can be analyzed with Monte-Carlo
methods.
 
In this article we present and study a version of the
Ising model which achieves this purpose. In Section 2 we
recall the basic features of Onsager's solution of the
model which employs a transfer matrix representation. We
review the diagonalization of the bulk transfer matrix in
terms of anticommuting spin matrices. In section 3
we turn to the issue of boundary conditions on an open
strip, and show that a simple condition on the original
Ising spins at the boundary produces the 1/2 integer
modes required of the Neveu-Schwarz field $H$. Section 4
deals with the explicit construction of the eigenoperators of the
transfer matrix. An important symmetry of the Neveu-Schwarz string
is the so-called G-parity. In Section 5 we identify the
symmetry of the Ising model that becomes G-parity in the
continuum limit. This is important for imposing the even
G-parity restriction in lattice simulations. Unfortunately,
imposing this restriction, in the most straightforward way, 
allows minus signs and/or nonlocality to creep back into 
the path summand, again posing potential difficulties for
Monte-Carlo methods. In Section 6, we discuss the 3 open string
vertex, which requires an operator insertion at the 
interaction point. It is argued that, provided the even G-parity
restriction is maintained, these insertion factors can
be taken into the exponent and interpreted as a modification of 
the worldsheet action. In Section 7 we give a brief discussion
of the resulting representation of the sum over planar diagrams.
Concluding discussion is in a final Section 8.

\section{Bulk properties of the Ising model}
We consider the two dimensional Ising model
on an $M\times N$ lattice specified by
the partition function
\bea
Z=\sum_{\sigma_i^j=\pm1}e^{\sum_{ij}(J 
\sigma_i^j\sigma_{i+1}^j+J^\prime\sigma_i^j\sigma_i^{j+1})/2}\;,
\label{isingpart}
\eea
where $1\leq i\leq M$ and $1\leq j\leq N$.
Onsager's transfer matrix representation of $Z$ is
\bea
Z&=&\left({e^{J^\prime}-e^{-J^\prime}}
\right)^{MN/2}\bra{f}{\cal T}(M)^N\ket{i}\label{onsagerpart}\\
\left({e^{J^\prime}-e^{-J^\prime}}
\right)^{M/2}{\cal T}(M)&=&\prod_k(e^{J^\prime/2}+\sigma_k^x
e^{-J^\prime/2})\exp\left\{{J\over2}
\sum_{i=1}^{M-1}\sigma_i^z\sigma_{i+1}^z\right\}
\nonumber\\
{\cal T}(M)&=&\exp\left\{{\xi\over2}\sum_{i=1}^M\sigma_i^x\right\}
\exp\left\{{J\over2}
\sum_{i=1}^{M-1}\sigma_i^z\sigma_{i+1}^z\right\}\label{tmatrix}\\
\tanh{\xi\over2}&=&e^{-J^\prime}\;,
\eea
where the states $\ket{i},\ket{f}$ belong to an $M$-fold tensor product
of 2-spinors, and they are determined by
the boundary conditions at $j=1,N$ respectively.
Here the $\sigma_i^{x,y,z}$ are $M$ independent
sets of $2\times2$ Pauli spin matrices:
\bea
\{\sigma_i^a,\sigma_i^b\}=2\delta_{ab},\qquad [\sigma_i^a,\sigma_j^b]
=0,\quad {\rm for}~i\neq j\; .
\eea
An important property of the eigenvalue spectrum of the
transfer matrix ${\cal T}(M)$ is an easy consequence of the
representation (\ref{tmatrix}): it is geometrically symmetric
about 1. To see this note that if ${\cal T}$ 
has the eigenvalue $T$ on the state $\ket{T}$, 
then it  has the
eigenvalue $T^{-1}$ on the state $e^{\xi\sum_k\sigma^x_k/2}
\prod_{k=\rm odd}\sigma^x_k\prod_l\sigma^z_l\ket{T}$. 
As shown by Onsager, the Ising model has a critical point
when $\xi=J$. For the isotropic case $J^\prime=J$, this occurs
when $\sinh J=1$, or $J=\ln(1+\sqrt{2})$.

As usual, it is most convenient to replace the Pauli
matrix dynamical variables with variables that 
anticommute for different $i$ using the Jordan-Wigner trick
\bea
S_i^{y,z}&\equiv&{\sigma_i^{y,z}\over\sqrt{2}}\prod_{k=1}^{i-1}\sigma_k^x\label{jordanwigner}\\
\{S_i^a,S_k^b\}&=&\delta_{ab}\delta_{ik}\;,
\eea
whereupon the Onsager transfer matrix becomes:
\bea
{\cal T}(M)&=& e^{-i\xi\sum_{k=1}^M S_k^y S_k^z}
e^{-iJ\sum_{k=1}^{M-1} S_{k+1}^z S_k^y}\; .\label{transfer}
\eea
This choice for ${\cal T}$ imposes a particular set of 
boundary conditions at $k=1,M$ appropriate for the Ising
spins living on an open strip. If they lived on a cylinder,
periodic or anti-periodic boundary conditions would be appropriate.
We shall show that the boundary conditions chosen here lead
to a half-integer moded fermion field in the continuum limit,
which is what is needed to describe the Neveu-Schwarz model
\cite{neveuschwarz,neveust}.
 
It is straightforward to calculate the action of ${\cal T}$
by conjugation on the $S$ variables:
\bea
{\cal T}S_k^z{\cal T}^{-1}&=& c_J c_\xi S_k^z
-ic_J s_\xi S_k^y +i s_J  c_\xi S_{k-1}^y
-s_J s_\xi S_{k-1}^z,\qquad 1<k\leq M\label{recursive1}\\
{\cal T}S_k^y{\cal T}^{-1}&=& c_J c_\xi S_k^y
+ic_J s_\xi S_k^z -i s_J  c_\xi S_{k+1}^z
-s_J s_\xi S_{k+1}^y,\qquad 1\leq k<M\label{recursive2}\\
{\cal T}S_1^z{\cal T}^{-1}&=&  c_\xi S_1^z
-i s_\xi S_1^y\label{recursive3} \\
{\cal T}S_{M}^y{\cal T}^{-1}&=&  c_\xi S_M^y
+i s_\xi S_M^z\; , 
\label{recursive4}
\eea
where we have introduced the shorthand notation $c_J\equiv\cosh J$,
$s_J\equiv\sinh J$, $t_J\equiv\tanh J$, and similarly for
$J\to\xi$.

We see that ${\cal T}$ acts linearly on the $S$'s, which
means that the problem of finding eigenoperators for
${\cal T}$ is one of linear algebra. The
form of the right sides of (\ref{recursive1}), (\ref{recursive2}) suggest
that an expansion of the $S$'s in eigenoperators will
involve plane wave $k$ dependence: 
\bea
S_k^z&=&\sum_\lambda A_\lambda e^{i\lambda k},\qquad 
S_k^y=\sum_\lambda B_\lambda e^{i\lambda k},\qquad
{\cal T}(A_\lambda,B_\lambda){\cal T}^{-1}=t_\lambda (A_\lambda,B_\lambda)\; .
\eea
We first consider the implications of the bulk recursion
relations, focusing on a particular mode $\lambda$. Suppressing
the $\lambda$ subscripts on $A,B,t$, we find
\bea
tA&=& c_J c_\xi A
-ic_J s_\xi B +i s_J  c_\xi Be^{-i\lambda}
-s_J s_\xi Ae^{-i\lambda} \nonumber\\
tB&=& c_J c_\xi B
+ic_J s_\xi A -i s_J  c_\xi Ae^{+i\lambda}
-s_J s_\xi Be^{+i\lambda}\; .\nonumber
\eea
The consistency of these two equations requires:
\bea
|t-c_J  c_\xi+e^{-i\lambda}s_J s_\xi|^2
=|c_J s_\xi-e^{-i\lambda}s_J c_\xi|^2\; ,
\eea
which simplifies to
\bea
t^2-2t(c_J c_\xi-s_J s_\xi \cos\lambda )+1=0\; .
\label{chpoly}
\eea
Since this equation is even under $\lambda\to-\lambda$ each 
of the two solutions, 
\bea
t_\pm=c_J c_\xi-s_J s_\xi \cos\lambda \mp\sqrt{
(c_J c_\xi-s_J s_\xi \cos\lambda )^2-1}\; ,
\label{toflambda}
\eea
belong to both left ($\lambda<0$) and right ($\lambda>0$)
moving plane waves. This degeneracy is important in 
realizing definite boundary conditions. Also from the equation,
it immediately follows that $t_+ t_-=1$. 

The eigenoperators $A_a$, with eigenvalue $t_a$, 
connect eigenstates of ${\cal T}$ 
with different eigenvalues.
Because the state space is of finite dimension there is
a maximum eigenvalue of ${\cal T}$, $T_{\rm max}$, and a minimum
eigenvalue: $T_{\rm min}=T_{\rm max}^{-1}$ by the symmetry
of the eigenvalue spectrum. Let $\ket{G}$ be the state
with the maximum eigenvalue. Then we must have
\bea
A_a\ket{G}=0,\qquad {\rm whenever}~ t_a >1\;,  
\eea
because if $t_a>1$, then the eigenvalue of the state on the
left would be $t_aT_{\rm max}>T_{\rm max}$. Assuming completeness
of the eigenoperators, i.e. that monomials of 
eigenoperators acting on $\ket{G}$ span the whole state space, 
then the the state $\prod_{t_a<1}A_a\ket{G}$ is the eigenstate with
the minimum eigenvalue $T_{\rm min}=T_{\rm max}\prod_{t_a<1}t_a$.
It follows that $T_{\rm max}^2=\prod_{t_a>1}t_a$:
\bea
T_{\rm max} = \sqrt{\prod_{t_a>1}t_a}\; .
\eea
Identifying $\ln T_{\rm max} = -E_G$ the ground state energy,
we can write this relation as
\bea
E_G= -{1\over 2}\sum_{t_a>1}\ln t_a=-{1\over 2}\sum_{t_a>1}\omega_a\; ,
\eea
where we have defined the energy created by an eigenoperator as
$\omega=-\ln t$. Then, since the eigenvalues $t$ come
in reciprocal pairs, the $\omega$'s come in opposite
sign pairs $\omega_-=-\omega_+$, and we
can write $\omega_\pm = \pm\omega$, with the convention that 
$\omega >0$. 

In order that infinitely long waves ($\lambda\to0$)
cost zero energy, we see from (\ref{toflambda}) that the Ising system must be
critical, that is $\xi=J$. In this critical case we can simplify
\bea
t_\pm&=&1+(1-\cos\lambda)s_J^2 \mp\sqrt{2(1-\cos\lambda)s_J^2 
+(1-\cos\lambda)^2s_J^4}\\
&=&1+2s_J^2\sin^2{\lambda\over2} \mp
2\left|s_J\sin{\lambda\over2}\right|\sqrt{1
+s_J^2\sin^2{\lambda\over2}}\\
&{\sim\atop{\lambda\to0}}&1 \mp
|{\lambda}s_J|\; .
\eea
Once $t$ is determined by consistency, we can then solve for $B$ in terms of 
$A$:
\bea
B={t-c_J c_\xi+e^{-i\lambda}s_J s_\xi\over
i(e^{-i\lambda}s_J c_\xi-c_J s_\xi)}A\equiv RA\; .
\eea
Notice that, according to the consistency condition, 
the modulus of the
numerator of $R$ is equal to the modulus
of the denominator, so $R$ is a pure phase.
Putting in the explicit forms for $t_\pm$, we get
\bea
R_\pm={-is_J s_\xi\sin\lambda \mp
\sqrt{(c_J c_\xi-s_J s_\xi\cos\lambda )^2-1}\over
i(e^{-i\lambda}s_J c_\xi-c_J s_\xi)}\; .
\label{bovera}
\eea
Specializing to the critical case, the form simplifies
\bea
R_\pm
=e^{i\lambda/2}{-is_J\cos{(\lambda/2)}\mp{\rm sgn}
(\lambda)
\sqrt{1+s_J^2\sin^2(\lambda/2)}\over c_J},\qquad \xi=J\; .
\label{rplusminus}
\eea
When we apply boundary conditions to an open chain in the
next section, 
it will be necessary
to consider linear combinations of left and right moving
plane waves. Then we will take $A,B$ as the coefficients of
$e^{i\lambda k}$ and $A^\prime, B^\prime$ as the coefficients
of $e^{-i\lambda k}$ with $\lambda>0$ in both cases. Then 
Eq.(\ref{bovera}) stands and
\bea
B^\prime&=& R^\prime A^\prime\\
R^\prime_\pm&=&{+is_J s_\xi\sin\lambda \mp
\sqrt{(c_J c_\xi-s_J s_\xi\cos\lambda )^2-1}\over
i(e^{+i\lambda}s_J c_\xi-c_J s_\xi)}=-R^*_\pm\; .
\label{boveraprime}
\eea
For the critical case these equations  become
\bea
R_\pm&=&e^{i\lambda/2}{-is_J\cos{(\lambda/2)}\mp
\sqrt{1+s_J^2\sin^2(\lambda/2)}\over c_J},\qquad \xi=J\\
R^\prime_\pm&=&
e^{-i\lambda/2}{-is_J\cos{(\lambda/2)}\pm
\sqrt{1+s_J^2\sin^2(\lambda/2)}\over c_J}=-R^*_\pm,
\qquad \xi=J\; .
\eea
\section{Boundary Conditions}
\begin{figure}[ht]
\begin{center}
\includegraphics[width=5in,height=4in]{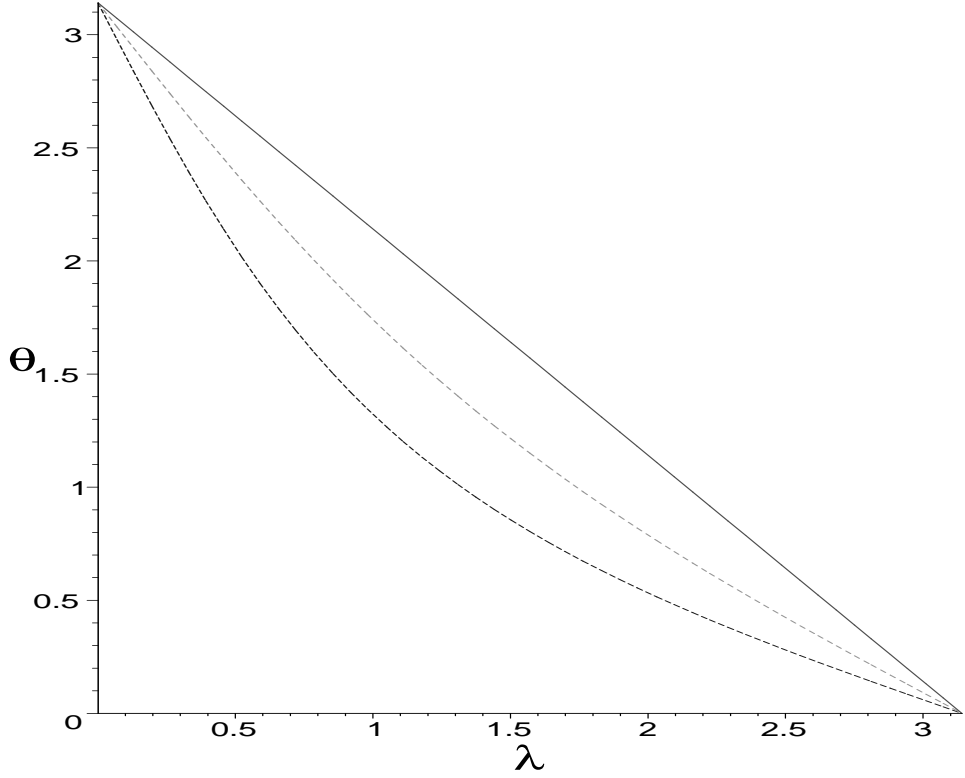}
\caption{$\theta(\lambda)$ for the critical case $\xi=J$, 
and $J=0$ (highest curve), $J=1$
(middle curve) and $J=1.5$ (lowest curve).}
\label{thetaoflambda}
\end{center}
\end{figure}

\begin{figure}[ht]
\begin{center}
\includegraphics[width=4.6in,height=3.6in]{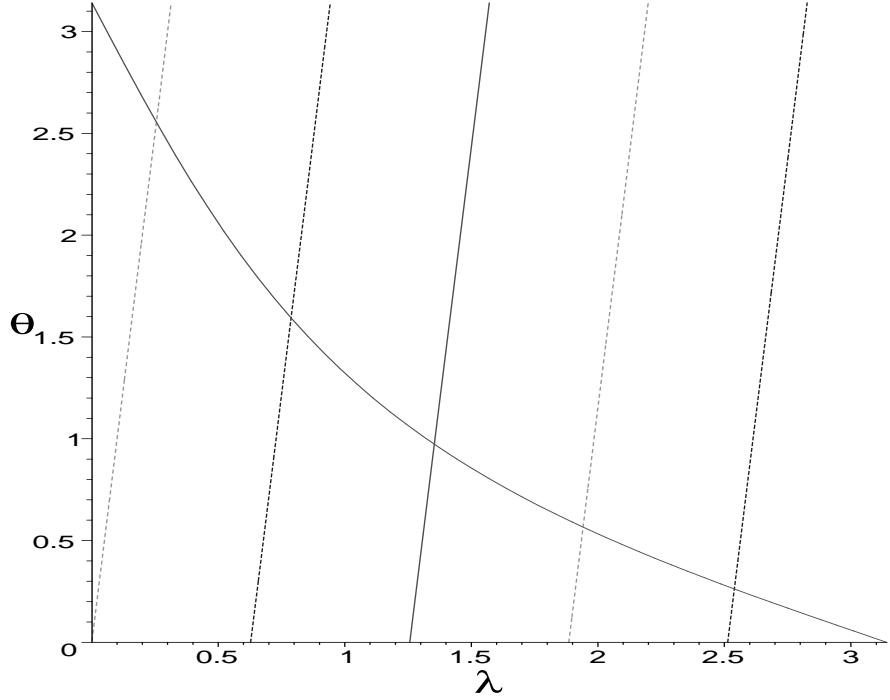}
\caption{Solution of (\ref{critlambda}) for $J=1.5$. The straight
lines are $2M\lambda-2n\pi$ for $n=0,1,\dots M-1$ for the case
$M=5$. Note that $\lambda=\pi$ corresponding to $n=M$, though a
solution of (\ref{critlambda}), is spurious as discussed in
the text.}
\label{lambdasolution}
\end{center}
\end{figure}

\begin{figure}[ht]
\begin{center}
\includegraphics[width=5in,height=4in]{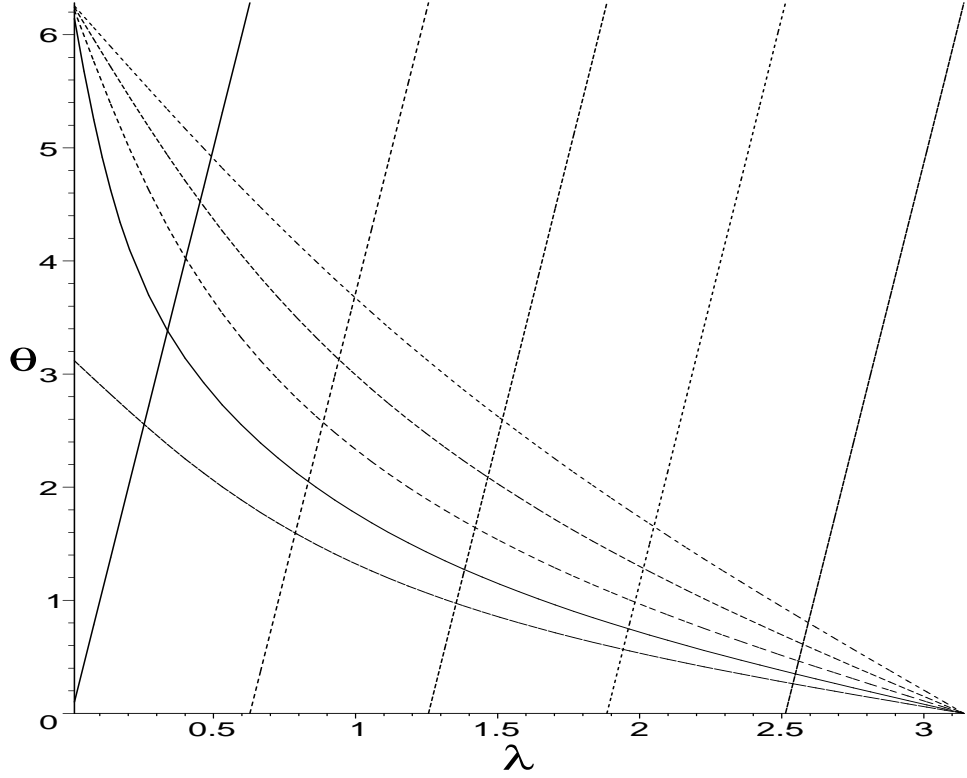}
\caption{$\theta(\lambda)$ for some noncritical cases: $\xi=1.5$ 
and $J=0.3,0.6,0.9,1.2$ from the highest to second lowest
curves. The lowest curve is the critical case $J=1.5$, 
included for comparison. The straight
lines are $2M\lambda-2n\pi$ for $n=0,1,\dots M-1$ for the case
$M=5$. Their intersections with the $\theta(\lambda)$ curves show
how the solutions of the eigenvalue equation for the noncritical
cases approach the solution for the critical case as $J\to\xi$.}
\label{thetaoflambdanoncrit}
\end{center}
\end{figure}

Now we turn to the issue of boundary conditions. 
The simplest open chain
model, given by the transfer matrix of (\ref{transfer}), will
suffice for our purposes. The action
of ${\cal T}$ on the first ($S_1^a$) and last 
($S_M^a$) spins (\ref{recursive3}), (\ref{recursive4}) 
must be made to agree with the
bulk recursive formulas (\ref{recursive1}), (\ref{recursive2}) 
which were satisfied with our plane wave ansatz. 
This can be arranged by suitable
definitions of $S_0^a$ and $S_{M+1}^a$, which do not appear in 
${\cal T}$. The bulk recursions read for $k=1,M$:
\bea
{\cal T}S_1^z{\cal T}^{-1}&=& c_J c_\xi S_1^z
-ic_J s_\xi S_1^y +i s_J  c_\xi S_{0}^y
-s_J s_\xi S_{0}^z \nonumber\\
{\cal T}S_M^y{\cal T}^{-1}&=& c_J c_\xi S_M^y
+ic_J s_\xi S_M^z -i s_J  c_\xi S_{M+1}^z
-s_J s_\xi S_{M+1}^y\; .\nonumber
\label{endrecursive}
\eea
Agreement will occur if we impose
\bea
(c_J-1)( c_\xi S_1^z
-i s_\xi S_1^y) + s_J (i c_\xi S_{0}^y
- s_\xi S_{0}^z) &=& 0\\
(c_J-1)( c_\xi S_M^y
+i s_\xi S_M^z) + s_J (-i c_\xi S_{M+1}^z
- s_\xi S_{M+1}^y)&=&0\; .
\eea
Then plugging the plane wave ansatz
\bea
S_k^z=A e^{i\lambda k}+A^\prime e^{-i\lambda k}, 
\qquad S_k^y=B e^{i\lambda k}+B^\prime e^{-i\lambda k}
\eea
into these equations and rearranging leads to
two different expressions for $\alpha$ defined
by $A^\prime=\alpha A$:
\bea
&&\hskip-.5in\alpha\ =\
-{e^{i\lambda}(c_J-1)( c_\xi -iRs_\xi) 
+ s_J( iR c_\xi 
- s_\xi)\over e^{-i\lambda}(c_J-1)( c_\xi 
-iR^\prime s_\xi) 
+ s_J(iR^\prime c_\xi - s_\xi)}\\
&&\hskip-.5in\alpha \ = -e^{2iM\lambda}
{(c_J-1)(Rc_\xi +i s_\xi) 
+ e^{i\lambda}s_J(-i c_\xi 
-Rs_\xi)\over(c_J-1)(R^\prime c_\xi 
+i s_\xi) 
- e^{-i\lambda}s_J(i c_\xi+ R^\prime s_\xi)}\; .
\eea
Remembering that $R$ and $R^\prime=-R^*$ have
already been determined in (\ref{bovera}), (\ref{boveraprime}), 
we see that these two equations
determine $\alpha$ and a quantization condition on
$\lambda$ from the consistency of the two equations.
We are particularly interested in the critical case $\xi=J$, 
for which the eigenvalue condition simplifies to
\bea
e^{2iM\lambda}={(e^{i\lambda}-1)c_J-e^{i\lambda}-1
\over(e^{i\lambda}-1)c_J+e^{i\lambda}+1}
=-\ {1-ic_J\tan{(\lambda/2)}
\over1+ic_J\tan{(\lambda/2)}}\equiv e^{i\theta(\lambda)}\; .
\label{critlambda}
\eea
We plot $\theta(\lambda)$ in the critical case for three different values
of $J=\xi$ in Fig.~\ref{thetaoflambda}.
The graphical solution of (\ref{critlambda}) is shown in 
Fig.~\ref{lambdasolution}, for $J=1.5$ and $M=5$. This graph
makes it clear that there are precisely $M$ nontrivial 
eigenoperator solutions for any value of $c_J$. The apparently
($M+1$)th solution $\lambda=\pi$ is spurious. 
This is because, in this
case, the left and right moving $k$ dependence is identical
($e^{ik\pi}=e^{-ik\pi}$) and the limit $\lambda\to\pi$
implies $C^\prime\to-C$ and $D^\prime\to-D$. Thus the
corresponding eigenoperators, proportional to $C+C^\prime$
and $D+D^\prime$, vanish.
For comparison, we show $\theta(\lambda)$ and the solution
of the eigenvalue equation for several noncritical values
$J<\xi$ in Fig.~\ref{thetaoflambdanoncrit}.

\begin{figure}[ht]
\begin{center}
\includegraphics[width=4in,height=3in]{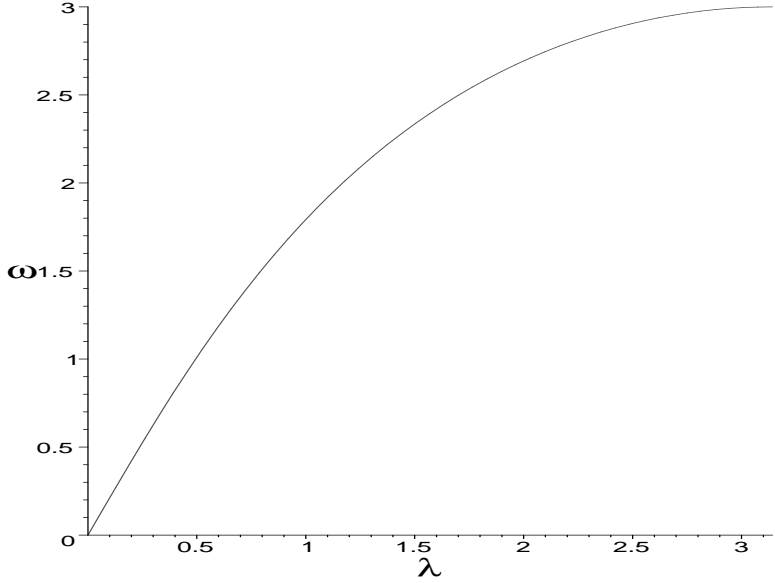}
\caption{The excitation energy $\omega(\lambda)=-\ln t_+(\lambda)$ 
for the critical case $\xi=J=1.5$.}
\label{omegaoflambdacrit}
\end{center}
\end{figure}

\begin{figure}[ht]
\begin{center}
\includegraphics[width=4in,height=3in]{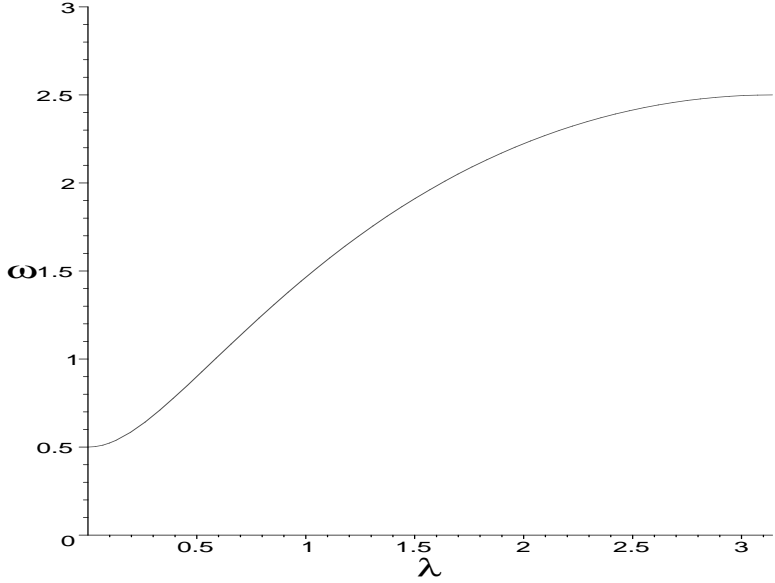}
\caption{The excitation energy $\omega(\lambda)=-\ln t_+(\lambda)$ 
for the noncritical case $J=1$, $\xi=1.5$.}
\label{omegaoflambdanoncrit}
\end{center}
\end{figure}

Applying an eigenoperator to an eigenstate of ${\cal T}$ with
eigenvalue $T_0$ produces
another eigenstate of ${\cal T}$ with eigenvalue $tT_0$. We can
say that the eigenoperator has increased the energy of the
state by an amount $\omega_\pm=-\ln t_\pm=\pm\omega$. 
We show $\omega(\lambda)$ for a critical case $\xi=J$ 
in Fig.~\ref{omegaoflambdacrit} and for a noncritical
case in Fig.~\ref{omegaoflambdanoncrit}.
The fact that the eigenvalue spectrum is non-degenerate 
guarantees that the set of eigenoperators is complete.

The critical case allows a nontrivial continuum limit $M\to\infty$: 
In that limit there are finite
energy excitations which have $\lambda=O(1/M)$. In this limit the
eigenvalue condition simplifies even further to
\bea
e^{i(2M+c_J)\lambda}=-1+O(\lambda^2),\qquad {\rm for}~ M\to\infty,~ 
\lambda M~ {\rm fixed}\; .
\eea
or $\lambda=(n+1/2)\pi/M$, which is the mode quantization of the
Neveu-Schwarz worldsheet field.

\section{Construction of Eigenoperators}
In this section we express the eigenoperators in
terms of the $S$'s.
We write the eigenoperator as
\bea
A=\sum_{k=1}^M(S_k^zU_k+S_k^yV_k),\qquad 
{\cal T}A{\cal T}^{-1}=tA\; .
\eea
Then we easily derive the recursion relations
satisfied by the $U,V$:
\bea
tU_k&=&
c_Jc_\xi U_k-s_Js_\xi U_{k+1}+ic_Js_\xi V_k-is_Jc_\xi V_{k-1},
\qquad k=2,\ldots,M-1\\
tV_k&=&
c_Jc_\xi V_k-s_Js_\xi V_{k-1}-ic_Js_\xi U_k+is_Jc_\xi U_{k+1},
\qquad k=2,\ldots,M-1\\
tU_1&=&c_\xi U_1-s_js_\xi U_2 +ic_Js_\xi V_1\\
tV_1&=&-is_\xi U_1+c_Jc_\xi V_1+is_Jc_\xi U_2\\
tU_M&=&c_Jc_\xi U_M-is_Jc_\xi V_{M-1}+is_\xi V_M\\
tV_M&=&c_\xi V_M-ic_Js_\xi U_M-s_Js_\xi V_{M-1}\; .
\eea
The last four special cases will be included in the
first two generic equations if we put
\bea
is_JV_0=(c_J-1)U_1,\qquad -is_JU_{M+1}=(c_J-1)V_M\;,
\eea
which specify the boundary conditions.
As before, these relations can be solved with the plane wave
ansatz:
\bea
U_k=Ce^{ik\lambda}+C^\prime e^{-ik\lambda},\qquad 
V_k=De^{ik\lambda}+D^\prime e^{-ik\lambda}\; ,
\eea 
leading to
\bea
{D\over C}&=&-i{t-c_Jc_\xi+e^{i\lambda}s_Js_\xi\over
c_Js_\xi-e^{-i\lambda}s_Jc_\xi}\\
{D^\prime\over C^\prime}&=&-{D^*\over C^*}\; ,
\eea
together with the consistency condition
\bea
t^2-2t(c_Jc_\xi-s_Js_\xi\cos\lambda)+1=0\; ,
\eea
which is identical to (\ref{chpoly}).
Note that $D/C$ is similar to $B/A$, but not identical to it.
This is because the matrix we are diagonalizing is not hermitian.
Of course, since the characteristic equation is the same for both
eigenvalue problems, the eigenvalue spectrum will be identical.
The boundary conditions read
\bea
is_J(D+D^\prime)&=&(c_J-1)(Ce^{i\lambda}+C^\prime e^{-i\lambda})\\
-is_J(Ce^{i(M+1)\lambda}+C^\prime e^{-i(M+1)\lambda})
&=&(c_J-1)(De^{iM\lambda}+D^\prime e^{-iM\lambda})\; ,
\eea
which lead to 
\bea
{C^\prime\over C}={is_J(D/C)-(c_J-1)e^{i\lambda}
\over is_J(D^*/C^*)+(c_J-1)e^{-i\lambda}}\; ,
\eea
and to the consistency condition
\bea
e^{2iM\lambda}&=&{(is_J(D/C)-(c_J-1)e^{i\lambda})
(-is_Je^{-i\lambda}+(c_J-1)(D^*/C^*))\over
(is_J(D^*/C^*)+(c_J-1)e^{-i\lambda})
((c_J-1)(D/C)+is_Je^{i\lambda})}\\
&=&{is_J(1+|D/C|^2)+2{\rm Re}(e^{-i\lambda}D/C)+2ic_J{\rm Im}(e^{-i\lambda}D/C)
\over is_J(1+|D/C|^2)-2{\rm Re}(e^{-i\lambda}D/C)
+2ic_J{\rm Im}(e^{-i\lambda}D/C)}\\
&=&{is_J+{\rm Re}(e^{-i\lambda}D/C)+ic_J{\rm Im}(e^{-i\lambda}D/C)
\over is_J-{\rm Re}(e^{-i\lambda}D/C)
+ic_J{\rm Im}(e^{-i\lambda}D/C)}\; .
\eea
The last line follows because one can show that $C/D$ is a pure phase.
Plugging the solution for $t$ in the equation for $D/C$ and 
a little manipulation shows that
\bea
e^{-i\lambda}{D\over C}&=&-i{\pm\sqrt{(c_Jc_\xi-s_Js_\xi\cos\lambda)^2-1}
+is_Js_\xi\sin\lambda\over c_Js_\xi\cos\lambda -s_Jc_\xi+ic_Js_\xi\sin\lambda}
\\
&\to&-{\pm\sqrt{1+s^2_j\sin^2(\lambda/2)}
+is_J\cos(\lambda/2)\over c_J(i\sin(\lambda/2)+\cos(\lambda/2))}\; ,
\eea
where the last line shows the simplifying critical limit $\xi\to J$.
In this limit we also evaluate
\bea
{\rm Re}\left[e^{-i\lambda}{D\over C}\right]&=&-{1\over c_J}\cos{\lambda\over2}
\left(\pm\sqrt{1+s^2_J\sin^2{\lambda\over2}}+s_J\sin{\lambda\over2}\right)\\
s_J+c_J{\rm Im}\left[e^{-i\lambda}{D\over C}\right]&=&
\sin{\lambda\over2}\left(\pm\sqrt{1+s^2_J\sin^2{\lambda\over2}}
+s_J\sin{\lambda\over2}\right)\; .
\eea
Then the quantization condition on $\lambda$ simplifies dramatically to
\bea
e^{2iM\lambda}
&=&-{\cos{(\lambda/2)}-ic_J\sin{(\lambda/2)}
\over \cos{(\lambda/2)}+ic_J\sin{(\lambda/2)}}\ =\ 
-{1-ic_J\tan{(\lambda/2)}
\over1+ic_J\tan{(\lambda/2)}}\; ,
\eea 
which is, of course, identical to (\ref{critlambda}).

\section{G-Parity on the Lattice}
In the Neveu-Schwarz model the G-parity
is a crucial symmetry concept. The fermionic raising and lowering
operators $b_r=b^\dagger_{-r}$ are moded with respect
to $R=\sum_{r>0}rb_{-r}br$ by half odd integers.
G-parity is defined by $G b_r G=-b_r$, and $G^2=1$. 
Defining fermion number as $N_F=\sum_{r>0}b_{-r}b_r$, 
$G$ is essentially $(-)^{N_F}$. Because of the circumstance
that the $b$'s carry half odd mode number, we could also
write $(-)^{N_F}=(-)^{2R}$ which leads to technical simplifications
in analyzing the model.
The importance of
the symmetry is that, since the interactions conserve
G-parity, it is consistent to restrict the open
string spectrum to be even under G-parity.
Since the lowest odd G-parity state is a tachyon,
and the lowest even G-parity state is a massless
vector, the restriction to even G-parity removes
the tachyon, leaving a massless sector that can
describe a gauge boson. With this convention for
even and odd, $G=-(-)^{N_F}=-(-)^{2R}$.

There is no lattice analog of ${N_F}$ that commutes with the
transfer matrix of the Ising model. 
However, there is a candidate for
$(-)^{N_F}$, namely 
$G=\eta\prod_k\sigma_k^x$, with $\eta^2=1$. This operator,
which anticommutes with $S_k^{y,z}$, commutes with the
transfer matrix. The Neveu-Schwarz model in $D$ space-times dimensions
requires $D-2$ transverse components ${\bfs H}$ and therefore $D-2$
Ising systems, describable by $D-2$ sets of Pauli matrices 
$\sigma^{x,y,z}_{iA}$, $A=1,\ldots,D-2$. Then the total 
G-parity operator will be $\eta\prod_{kA}\sigma^x_{kA}$.
It remains to fix the value of
$\eta=\pm1$. According to the conventions of the
Neveu-Schwarz model, we would like it to have the
value $-1$ on the ground state of the system 
when it is critical. Since the ground state at general
$J$ has only been determined implicitly by rather complicated
equations, it would seem a daunting task to evaluate the
action of $G$ on it. However, because $G$ takes on 
only the values $\pm1$, its value cannot change under the
variation of a continuous parameter. Thus we can exploit
simplifications in the state that occur when $J\to0$
at fixed $\xi$ and $M$.

As we can see from Fig.~\ref{thetaoflambdanoncrit}, the
eigenvalue problem varies continuously as $J$ decreases
continuously from the $J=\xi$ to $J=0$. In the limit $J\to0$
the eigenvalue equation smoothly approaches
\bea
e^{2iM\lambda}&=&e^{-2i\lambda},\quad 
\Rightarrow \lambda={n\pi\over M+1},
\qquad n=1,\cdots, M\; .
\eea
Furthermore the eigenvalues $t_\pm(\lambda)\to e^{\mp\xi}$,
independent of lambda, and the eigenoperators approach
\bea
A_n^\pm\to 2iC\sum_{k=1}^M(S_k^z\pm iS_k^y)\sin{n\pi k\over M+1}\; .
\eea
The $A_n^-$ increase the eigenvalue of a state by the
factor $e^\xi$.
Therefore in the limit the ground state of the system must
be annihilated by all the $A_n^-$. Since the mode functions
$\sin(n\pi k/(M+1))$ are complete, this limiting state must
satisfy
\bea
(S_k^z-iS_k^y)\ket{G_0}=0,\qquad {\rm for~all}~ k=1,\cdots, M\; .
\eea
But $\sigma_k^x=i\sigma_k^z\sigma_k^y=2iS_k^zS_k^y$ so
it follows that
\bea
\sigma_k^x\ket{G_0}&=&2iS_k^z(-iS_k^z)\ket{G_0}
=\ket{G_0},\qquad {\rm for~all}~k\\
\prod_k\sigma^x_k\ket{G_0}&=&+\ket{G_0}\; .
\eea
By continuity we conclude that the ground state
at all $J<\xi$ will have this same eigenvalue.
This conclusion depends on the equations determining
the ground state not changing discontinuously at some point.
For example if the eigenvalue of an eigenoperator changed
from $t>1$ to $t<1$ (from $\omega<0$ to $\omega>0$) as $J$ was
varied, it would no longer
annihilate the ground state and vice versa. But the
character of the eigenvalue solutions doesn't
allow this to happen, as is evident from 
Fig.~\ref{thetaoflambdanoncrit}. In particular note that the
curve in Fig.~\ref{omegaoflambdanoncrit} stays well
away from $\omega=0$ throughout.
We conclude that $\eta=-1$, so the correct G-parity operator to use
in describing the Neveu-Schwarz model is
$G=-\prod_{kA}\sigma^x_{kA}$.

Finally, we work out the meaning of G-parity in the
language of the original Ising partition function (\ref{isingpart}).
To do this, focussing on a single Ising system,
we consider the product of the appropriate factors of 
$G$ with the transfer matrix (\ref{tmatrix}):
\bea
\prod_k\sigma_k^x{\cal T}
&=&(e^{J^\prime}-e^{-J^\prime})^{-M/2}
\prod_k\left(e^{-J^\prime/2}+\sigma_k^x e^{J^\prime/2}\right)
\exp\left\{{J\over2}\sum_k\sigma_k^z\sigma_{k+1}^z\right\}\;,
\eea
compared to
\bea
{\cal T}&=&(e^{J^\prime}-e^{-J^\prime})^{-M/2}
\prod_k\left(e^{J^\prime/2}+\sigma_k^x e^{-J^\prime/2}\right)
\exp\left\{{J\over2}\sum_k\sigma_k^z\sigma_{k+1}^z\right\}\; .
\eea
Thus, inserting a $G$ somewhere in the 
matrix element (\ref{onsagerpart}) has the 
effect on (\ref{isingpart}) of multiplying
$Z$ by an overall factor $(-1)$ and reversing the sign of
{\it one} of the terms in the sum over $j$, 
$J^\prime\sum_j(\sum_{kA}\sigma_{kA}^j
\sigma_{kA}^{j+1})/2$. Technically this is accomplished by
inserting a factor $-e^{-J^\prime\sum_{kA}\sigma_{kA}^l
\sigma_{kA}^{l+1}}$, for a particular time slice $l$, 
in the summand of (\ref{isingpart}).
To project onto even G-parity states, simply insert the
factor $(1-e^{-J^\prime\sum_{kA}\sigma_{kA}^l\sigma_{kA}^{l+1}})/2$.
Or perhaps a better way to say it is to make the
replacement 
\bea
e^{J^\prime\sum_{kA}\sigma_{kA}^l\sigma_{kA}^{l+1}/2}
\to \sinh\left\{{J^\prime\over2}\sum_{kA}\sigma_{kA}^l\sigma_{kA}^{l+1}
\right\}\eea
on at least one time slice of each open string propagator. This
sinh factor is non local and also can be negative. The nonlocality
would add to the computational cost of a Monte Carlo simulation.
This is because the probabilistic criterion for
retaining an update of the spin on a site $k,l$ requires knowledge
of all the other sites on the time slice $l$, as well as those on neighboring
time slices. However, the one-dimensionality of this nonlocality
may help keep the cost manageable. The strict probabilistic interpretation
of the summand of $Z$, which is the theoretical basis
for Monte Carlo simulations, is marred by the fact that the sinh is
negative for some spin configurations. However the spin configurations
where the argument of the sinh is negative are strongly suppressed
by the exponential factors on the many other time slices. One
can then hope that those configurations will cause a minimal
degradation of the convergence of the simulation.  

\section{Open String Vertex}
\begin{figure}[ht]
\begin{center}
\includegraphics[width=2.3in]{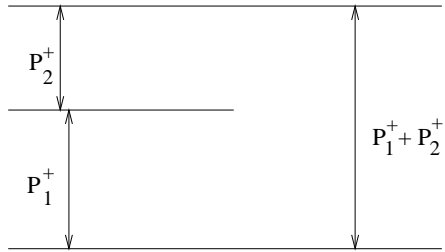}
\caption{Lightcone parameter domain for a three open string
vertex. Lightcone time $\tau$ is the horizontal axis and
$\sigma$ is the vertical axis.}
\label{lcvertex}
\end{center}
\end{figure}

A striking aspect of string theory is that
interactions among strings are inherent in the nature
of free string. This is because a single string can make
a transition to two strings by simply breaking at 
a point. 
Using light-cone quantization of the
free string \cite{goddardgrt} Mandelstam's interacting string
formalism \cite{mandelstamlc} provides the most concrete 
realization of this concept. 
As illustrated in 
Fig.~\ref{lcvertex}, Mandelstam's  three string vertex is simply a
worldsheet path integral with the free string action,
but for which the worldsheet fields
live on a two dimensional domain corresponding to 
two strings joining ends to become a single string (or the
time reversed process).
Discretizing this domain on a worldsheet grid, Giles and
I \cite{gilest} clarified the nature of the singularity induced
by the string joining/splitting process for the bosonic string
in $D$ space-time dimensions. Write $T=aN$, $P^+_1+P^+_2
\equiv P^+=MaT_0$, so the diagram has dimensions  $P^+\times T$,
and the associated lattice is $M\times N$. We found that,
in the continuum limit $M,N\to\infty$, the $360^\circ$ corner
at the interaction point induced a behavior $M^{-(D-2)/16}
\times$Finite. For the critical dimension, $D=26$, this
scaling behavior accounts precisely for the $(P^+_1P^+_2P^+)^{-1/2}
=(MaT_0)^{-3/2}(P_1^+P_2^+/P^{+2})^{-1/2}$ factor required
by Poincar\'e invariance.

For the Neveu-Schwarz model, a fermionic worldsheet field,
$H^\mu(z)=\sum_r b^\mu_rz^{-r}$ with $r$ ranging over 
half odd integers, is introduced. Of course, only the
transverse components $H^k$, $k=1,\ldots, D-2$ play
a role on the lightcone worldsheet. The main thrust of this
article is to represent these fields by $D-2$
independent critical Ising models on the worldsheet lattice. 
The contribution of the Ising degrees of freedom to the
singular behavior at the interaction point can be
inferred by realizing that a Majorana fermion is,
roughly speaking, a half boson. More precisely,
two Majorana fermion worldsheet fields can be interpreted,
through bosonization, as a single bosonic worldsheet field.
Thus the singular factor, including Ising and coordinate
variables, should be $M^{-3(D-2)/32}$. If this were the end of
the story, the Lorentz invariant critical dimension would
be $D=16+2=18$, not the Neveu-Schwarz critical dimension $D=10$.

The point is that the Neveu-Schwarz vertex is not simply the overlap
represented by the diagram of Fig.~\ref{lcvertex}, but there
is also an operator insertion at the joining point
\cite{mandelstamnsr}. In the continuum limit the
insertion is just the density of the superconformal
generators, ${\bfs H}\cdot{\dot{\bfs x}}$. Mandelstam
showed that if this insertion is placed a distance $\epsilon$
from the interaction point, then the amplitude $\sim
\epsilon^{-3/4}$ as $\epsilon\to0$. More generally,
if the operator insertion had conformal weight $J$, the singular 
behavior would be $\epsilon^{-J/2}$.

We now translate these conclusions to a lattice calculation.
Then there is no need to introduce $\epsilon$: one simply
inserts the operator a lattice step or two away from the
interaction site. Next, with lattice normalization
the insertion operator itself would correspond to
$a^J$ times the continuum expression. For example,
instead of ${\dot{\bfs x}}$, one would insert
${\bfs x}_i^{j+1}-{\bfs x}_i^j\to a{\dot{\bfs x}} $.
Likewise, instead of ${\bfs H}$ which has delta function
normalization, one would insert ${\bfs S}_i^j\to
a^{1/2}{\bfs H}$, which has Kronecker delta 
normalization. Thus putting $\epsilon=a$ the
continuum analysis of Mandelstam would lead to the
behavior $a^J\times a^{-J/2}=a^{J/2}$. Finally, in the
lattice setup there is no reference to $a$, only to the
number of lattice sites. Thus this estimate translates
to the behavior $1/M^{J/2}$ or $1/M^{3/4}$ for the
case $J=3/2$ of interest here. Putting this together with
the result from the overlap leads to the
overall scaling behavior 
\bea
V_{\rm NS}\sim \left({1\over M}\right)^{3(D-2)/32+3/4}
\eea
which is seen to give the correct Lorentz invariance
power $3/2$ for $D=8+2=10$ in accord with the known
properties of the model. This result has been obtained
by translating the analysis done in the continuum theory
into expectations for the results of a lattice calculation. It would be
very nice to also see it from a direct lattice calculation,
but we do not attempt that here.

Now let us look more closely at the operator insertion
from the Ising model point of view. We first need to
decide which Ising model will describe 
the $D-2\equiv d$ fermi fields ${S}_{kA}^{y,z}$, $A=1,\ldots,d$. 
The most straightforward
choice would be simply $d$ decoupled Ising models
as defined by (\ref{onsagerpart}), with each $S_{kA}^{y,z}$
built from the Pauli matrices following (\ref{jordanwigner}).
Then to make the $S_{kA}$ for different $A$ anticommute,
one could define individual G-parity operators $G_A=
\prod_{k=1}^M\sigma_{kA}^x$, and include a factor of
$\prod_{B=1}^{A-1}G_B$ in the definition of $S_{kA}$:
\bea
S^{y,z}_{kA}=\sigma_{kA}^{y,z}
\prod_{B=1}^{A-1}G_B \prod_{l=1}^{k-1}\sigma_{lA}^x,
\qquad {\rm JW~ I}\; .
\eea
The trouble with this choice is that the insertion operator
${\dot x}_A S_A$ will have a residual nonlocality when expressed
in terms of the original Ising variables.

There is a better choice, which can be described as follows.
Extend the index of $S^{y,z}_k$ to the range $k=1,\ldots, Md$.
Then identify the first $d$ of these with $S_{1A}$, the next
$d$ with $S_{2A}$, etc. Then relate the anticommuting $S_k^{y,z}$
to the Ising $\sigma_k^{y,z}$ by the standard Jordan-Wigner
transform (\ref{jordanwigner}):
\bea
S^{y,z}_k=\sigma^{y,z}_k\prod_{l=1}^{k-1}\sigma_l^x,\quad k=1,\cdots, Md
\qquad {\rm JW~II}\; .
\eea
The advantage of this version of the Jordan-Wigner trick is that
the nonlocality of the insertion operator is subsumed in 
a factor which is proportional to the G-parity operator
for one of the strings entering the vertex. When all strings 
entering the vertex are
restricted to even G-parity, the nonlocality in the Ising variables
disappears! Thus the insertion will be local in {\it both} the 
Pauli matrix and fermionic representations.
This is very welcome, because it is the Pauli matrix form
that will be more amenable to numerical analysis.
On the unrestricted state space 
it is only the fermionic representation that is local.

There is a price for this choice however. Since we demand that
the transfer matrix expressed in terms of the fermionic spin 
variables be unchanged, the new Jordan-Wigner transform leads
to a more complicated Ising model. To see how we plug the 
new Jordan-Wigner transform into the transfer matrix
\bea
{\cal T}(M)&=& e^{-i\xi\sum_{A=1}^{d}\sum_{k=1}^M S_{kA}^y S_{kA}^z}
e^{-iJ\sum_{A=1}^{d}\sum_{k=1}^{M-1} S_{k+1,A}^z S_{kA}^y}\nonumber\\
&=& e^{-i\xi\sum_{k=1}^{Md} S_{k}^y S_{k}^z}
e^{-iJ\sum_{k=1}^{(M-1)d} S_{k+d}^z S_{k}^y}\; .\nonumber\\
\eea
The terms in the exponent of the first factor behave exactly as before:
\bea
-i\sum_{k=1}^{Md} S_{k}^y S_{k}^z &=&-{i\over2}
\sum_{k=1}^{Md} \sigma_{k}^y \sigma_{k}^z ={1\over2} \sum_{k=1}^{Md}
\sigma_k^x\nonumber\\
(e^{J^\prime}-e^{-J^\prime})^{Md/2} 
e^{-i\xi\sum_{k=1}^{Md} S_{k}^y S_{k}^z}&=&
\prod_{k=1}^{Md}(e^{J^\prime}+\sigma_k^x e^{-J^\prime})\; .
\eea
However the exponent of the second factor changes substantially:
\bea
-iJ\sum_{k=1}^{(M-1)d} S_{k+d}^z S_{k}^y&=&
{J\over2}\sum_{k=1}^{(M-1)d}\sigma^z_{k+d}
\sigma^z_k\prod_{l=k+1}^{k+D-3}
\sigma_l^x\nonumber\\
(e^{J^\prime}-e^{-J^\prime})^{Md/2}{\cal T}(M)
&=&
\prod_{k=1}^{Md}(e^{J^\prime/2}+\sigma_k^x e^{-J^\prime/2})
\exp\left\{{J\over2}\sum_{k=1}^{(M-1)d}\sigma^z_{k+d}
\sigma^z_k\prod_{l=k+1}^{k+d-1}
\sigma_l^x\right\}\; .
\label{newtransfer}
\eea
Only for $d=1$ does this reduce to the usual Ising model.
The form of the Ising model corresponding to this transfer matrix
is worked out in the Appendix.

\section{Summing Planar Loops}
\begin{figure}[ht]
\begin{center}
\includegraphics[width=4in,height=2.2in]{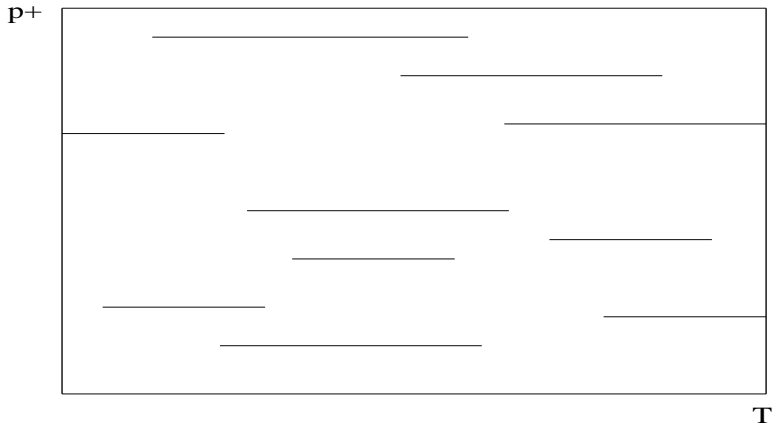}
\caption{A planar multiloop lightcone interacting string diagram. 
The horizontal lines form the boundaries of the propagator worldsheets
for intermediate open strings. This diagram has 7 loops and 5
external strings.}
\label{multiloop}
\end{center}
\end{figure}

Once the three open string vertex has been established
as in the previous section, the complete perturbation 
series is determined \cite{mandelstamlc,mandelstamnsr}. 
A generic planar multiloop diagram
in the series is the worldsheet path integral
using the free open string worldsheet action, but with the worldsheet
variables living on a domain such as depicted in 
Fig.~\ref{multiloop}.

In order to digitize the sum over planar diagrams, we fix the
overall dimensions of the domain to $P^+\times T$, and then
set up a worldsheet grid of dimensions $M\times N$, with
$T=aN$ and $P^+=aT_0 M$ \cite{gilest}. The discretized worldsheet
action is constructed so that the internal horizontal
lines shown in Fig.~\ref{multiloop} represent open string
boundaries. For the Neveu-Schwarz model the worldsheet
variables are the $d=D-2$ transverse coordinates ${\bfs x}(\sigma,\tau)
\to {\bfs x}_i^j$ and the $d$ fermionic fields 
$\sqrt{a}{\bfs H}(\sigma,\tau)\to {\bfs S}_i^j$ where the ${\bfs S}_i^j$
are taken to be the Jordan-Wigner transformed Pauli
spin matrices of $d$ independent Ising models. 
For each end of a horizontal line,
which depicts the breaking or joining of open strings,
there is a factor of the open string coupling $g$ and
also the operator insertion ${\cal S}_i^j\cdot
({\bfs x}_i^{j+1}-{\bfs x}_i^j)$ as explained in the
previous section. The precise location of this
insertion is somewhat flexible, as long as it
is within one or two lattice steps from the end of the 
horizontal line. For definiteness, we will
always place it on the longest string participating in
the vertex, with $i$
the spatial location of the horizontal line, and $j-1$
or $j+2$ the time of the end of the horizontal line,
with the choice determined so that the insertion
lies completely on the longest string participating in the
vertex. 

Next we turn to G-parity restrictions. The vertex
obtained in the previous section, and adopted in
this section, conserves G-parity: it connects 3
even G-parity open strings with each other or
2 odd G-parity strings to an even one. We would like to
restrict the open string states to even G-parity only.
When a diagram involves one or more loops, it is
not sufficient to restrict the external states to
even G-parity, because a pair of odd G-Parity
states can be produced from an even G-parity state.
Thus each internal propagator in a multiloop diagram
such as depicted in Fig.~\ref{multiloop}
must include a projector $(1+G)/2$ onto the even G-parity
sector.

As we have discussed in the previous section, 
the presence of even G-parity projectors throughout the
diagram introduces nonlocality into the worldsheet dynamics.
The projectors also produce negative contributions to
the path integrand.
These could potentially lead to inefficiencies and inaccuracies in
the applications of Monte Carlo algorithms to this system. 
However, there are beneficial aspects of the presence of the projectors. 
One, already mentioned,
is that restriction to the even G-parity states 
renders the nonlocality in the relation between
the $S$'s and the $\sigma$'s harmless. Thus the operator
insertions, necessary to describe the Neveu-Schwarz model,
will have a local representation in terms of the
original Ising variables. Even so, the insertions are
awkward because they are indefinite in sign and
have no obvious interpretation 
as terms in the worldsheet action.

It would be nice if these factors could be taken into the exponent where
they would become a modification of the action\footnote{
When ${\bfs H}$ is represented as a Grassmann variable this 
procedure would have the bizarre effect of adding 
Grassmann odd terms to the action. However in the Ising
model representation it would simply mean adding terms linear
in the spin to the action--no weirder than an external magnetic field!}.
If we do that and then expand the exponential,
the effect would be to replace the desired insertion 
with a sum of all possible powers of the insertion,
including a term with no insertion at all. The higher
powers are innocuous because they will either renormalize the
zeroth and first powers or introduce operators of 
higher conformal weight
which would be suppressed relative to the zeroth and
first powers in the continuum limit. It is
the zeroth term that poses the difficulty.
In the continuum limit it would dominate over the desired 
linear term.
On the other hand it does not couple 3 even G-parity
open strings together. Thus, if the even G-parity
restriction is enforced throughout, this troublesome
term will not contribute. So by including the projectors,
we enable the interpretation of the operator insertions
as modifications of the action. 

\section{Concluding Remarks}
In this article we have extended the GT lattice lightcone string
formalism to include the Neveu-Schwarz model, representing the
Fermi fields via Ising spin systems. That representation
has the virtue of avoiding the minus signs inherent in
a description of fermions as Grassmann fields. If there
were no need to include operator insertions at the interaction points
and no need to make the even G-parity restriction, the formalism
would be ready for immediate analysis via Monte Carlo simulations.

Alas, that would give neither the Neveu-Schwarz (NS) model nor
its even G-parity projected descendent (NS+). Indeed, the simulations
would be dominated by tachyon effects, most likely 
yielding the same sort of almost trivial outcome as the bosonic string
\cite{orland}, namely the sum of planar open string diagrams
would produce the propagator of a free closed string. What makes the
NS+ model dynamically interesting is precisely what makes
its Monte Carlo analysis problematic: the operator insertions
in the NS model are of indefinite sign and the projectors necessary
to describe the NS+ model are nonlocal. Intriguingly, for
the NS+ model, we argued that the restriction to even
G-parity provides a resolution to the operator insertion
difficulty. In its present status, all of the sign 
and nonlocality problems of the formalism 
reside entirely in the projection procedure. 
We think it is some progress to be able to
attribute all of the nonlocality and negativity to
such a well-defined source, in which they may turn well out
to be relatively benign. In any case, we suspect that
a better formulations will be found to circumvent 
any difficulties that remain.

Turning to more mundane matters, we recall that our main motivation
for keeping $\alpha^\prime>0$ was to mitigate the need for
uncontrolled counterterms to maintain Lorentz invariance in
the loop expansion. While this should effectively deal with the
usual field theoretic UV divergences, it does not by itself
take care of the potential need for the worldsheet contact
counterterms studied, for example, in \cite{greenseiberg}.
This issue clearly needs further study. However it is comforting
to know that the fundamentally stringy regularization identified in
\cite{GNS} (GNS), seems to be at least partially realized by the
GT lightcone lattice. The GNS mechanism, works by regarding
 each open string loop on the worldsheet as an emission or
absorption of a closed string in the vacuum: giving that
closed string a nonzero momentum regulates the worldsheet
divergence. By studying the one loop open string self energy,
we have found that discretizing $P^+=Ma$, which is the discretization
of GT lightcone lattice, has a similar regulating effect as $p\neq0$. 
For this example it guarantees that the open string gluon
remains massless! If this continues to happen in more
complicated multi-loop processes, we will be in business.

\vskip14pt

\noindent\underline{Acknowledgments}: 
I would like to acknowledge the
hospitality of the School of Natural Sciences
at the Institute for Advanced Study, where the early stages of this
work began.
This research was supported in part by the Ambrose Monell
Foundation and in part by the Department
of Energy under Grant No. DE-FG02-97ER-41029.

\appendix
\section{Alternate Description of $d$ Ising Spin Systems}
Here we work out the Ising model corresponding to the transfer matrix
(\ref{newtransfer}). The way to do this is to evaluate the
transfer matrix elements between basis states which are eigenstates of
the $\sigma_k^z$. Then ${\cal T}^N$ can be expressed as sums over
these eigenvalues. However it is more convenient to factor
${\cal T}$ into $d=D-2$ factors, and work out the matrix elements of each
factor. So write
\bea
(e^{J^\prime}-e^{-J^\prime})^{Md/2}{\cal T}(M)
&=&\prod_{m=1}^{d}\Bigg[\prod_{k=1}^{M}(e^{J^\prime/2}
+\sigma_{m+(k-1)d}^x e^{-J^\prime/2})
\nonumber\\&&\hskip.4in
\prod_{k=1}^{M-1}\exp\Bigg\{{J\over2}\sigma^z_{m+kd}
\sigma^z_{m+(k-1)d}\prod_{l=m+1+(k-1)d}^{m-1+kd}
\sigma_l^x\Bigg\}\Bigg]\; ,
\eea
and we evaluate the matrix elements of each $T_m$, where
\bea
T_m&=&\prod_{k=1}^{M}(e^{J^\prime/2}+\sigma_{m+(k-1)d}^x e^{-J^\prime/2})
\prod_{k=1}^{M-1}\exp\Bigg\{{J\over2}\sigma^z_{m+kd}
\sigma^z_{m+(k-1)d}\prod_{l=m+1+(k-1)d}^{m-1+kd}
\sigma_l^x\Bigg\}\nonumber\\
&=&\prod_{k=1}^{M}(e^{J^\prime/2}+\sigma_{m+(k-1)d}^x e^{-J^\prime/2})
\nonumber\\&&\hskip.5in
\prod_{k=1}^{M-1}\Bigg\{\cosh{J\over2}+\sigma^z_{m+kd}
\sigma^z_{m+(k-1)d}\prod_{l=m+1+(k-1)d}^{m-1+kd}
\sigma_l^x\sinh{J\over2}\Bigg\}\; .
\eea
Notice that in the last expression no $\sigma^x$ appears more than linearly.
Thus we can use the identities
\bea
\bra{\sigma^\prime}\sigma^x\ket{\sigma}&=&{1-\sigma^\prime\sigma\over2},
\qquad \bra{\sigma^\prime}I\ket{\sigma}={1+\sigma^\prime\sigma\over2}\; ,
\eea
where $\ket{\sigma}$ is an eigenstate of $\sigma^z$ with eigenvalue
$\sigma$. Then a few minutes thought leads to 
\bea
\bra{\{\sigma^\prime\}}T_m\ket{\{\sigma\}}&=& \exp\Bigg\{
{J^\prime\over2}\sum_{k=1}^M\sigma^\prime_{m+(k-1)d}
\sigma_{m+(k-1)d}\Bigg\}\nonumber\\
&&\hskip24pt\prod_{k=1}^{M-1}\Bigg[\prod_{l=m+1+(k-1)d}^{m-1+kd}
{1+\sigma^\prime_l\sigma_l\over2}\cosh{J\over2}
\nonumber\\&&\hskip54pt
+\sigma_{m+kd}\sigma_{m+(k-1)d}
\prod_{l=m+1+(k-1)d}^{m-1+kd}{1-\sigma^\prime_l\sigma_l\over2}
\sinh{J\over2}\Bigg]\; .
\eea
The matrix element of the full transfer matrix is then
\bea
\bra{\{\sigma^\prime\}}\prod_{m=1}^{d}T_m\ket{\{\sigma\}}
&=&\sum_{\{\{\sigma_i\}\}}\bra{\{\sigma^\prime\}}T_1\ket{\{\sigma_1\}}
\bra{\{\sigma_1\}}T_2\ket{\{\sigma_2\}}\cdots
\bra{\{\sigma_{D-3}\}}T_{d}\ket{\{\sigma\}}\; .
\eea
In effect we can think of each actual time slice as $d$ coincident
time slices, so that the computational scale of this new version
of the Ising model would correspond to an $M\times dN$ lattice
as compared to an $M\times N$ grid of the simple Ising model.
And of course as complicated as this formulation seems, it
has identical physics to $d$ decoupled Ising models. The only reason
for suffering these extra complications is that in the application to
the interacting NS+ model, the insertion operators will be local
in the new representation, whereas they would be nonlocal in
the original formulation.

\end{document}